\documentclass{ws-ijmpa}

\newcommand{\lt}{\left}
\newcommand{\rt}{\right}
\newcommand{\no}{\nonumber}
\newcommand{\nn}{\nonumber \\}
\newcommand{\ov}[1]{\overline{#1}}
\newcommand{\eq}[1]{Eq.~(\ref{#1})}

\newcommand{\imag}{\mathrm{Im}\,}
\newcommand{\real}{\mathrm{Re}\,}

\newcommand{\sgn}{\mbox{sign}\,}

\newcommand{\Bbar}{\,\overline{\rm B}}

\newcommand{\bbd}{\ensuremath{\rm B_d\!-\!\Bbar{}_d\,}}
\newcommand{\bbs}{\ensuremath{\rm B_s\!-\!\Bbar{}_s\,}}

\newcommand{\bbms}{\bbs\ mixing}
\newcommand{\bbmd}{\bbd\ mixing}
\newcommand{\BsorBsbar}{\raisebox{7.7pt}{$\scriptscriptstyle(\hspace*{8.5pt})$}
  \hspace*{-10.7pt}\!\Bbar_{s}}
\newcommand{\nuornubar}{
  \,\raisebox{5.5pt}{$\scriptscriptstyle(\hspace*{6.3pt})$}
  \hspace*{-7.8pt}\!\ov{\nu}}
\newcommand{\bra}[1]{\ensuremath{\langle #1 |}}
\newcommand{\ket}[1]{\ensuremath{| #1 \rangle }}
\newcommand{\fig}[1]{Fig.~\ref{#1}}

\newcommand{\lbar}{\ov{\Lambda}}
\newcommand{\dm}{\ensuremath{\Delta M}}
\newcommand{\dg}{\ensuremath{\Delta \Gamma}}

\newcommand{\epm}[2]{
 \raisebox{-0.5ex}{\shortstack[l]{$\scriptstyle+#1$\\$\scriptstyle-#2$}}}

\begin{document}

\markboth{Ulrich Nierste}
{Bounds on new physics from $\rm B_{\rm s}$ mixing}

%
\catchline{}{}{}{}{}
%

\title{TTP07-18\hfill arXiv:yymm.nnnn [hep-ph]\\[5mm]
Bounds on new physics from $\rm B_{\rm s}$ 
mixing\footnote{in collaboration with Alexander Lenz. Talk at 
\emph{CTP Symposium On Supersymmetry At LHC: Theoretical And
  Experimental Prospectives}, 11-14 Mar 2007, Cairo, Egypt.}
}

\author{Ulrich Nierste
}

\address{
        Institut f\"ur Theoretische Teilchenphysik (TTP)\\
        Karlsruhe Institute of Technology --- Universit\"at Karlsruhe\\
        76128 Karlsruhe\\
        Germany\\
        nierste@particle.uni-karlsruhe.de}

\maketitle

\begin{history}
\received{}
\revised{}
\end{history}

\begin{abstract}
  I summarize the observables constraining the \bbms\ complex and
  present a new calculation of the element $\Gamma_{12}^s$ of the 
  decay matrix. $\Gamma_{12}^s$ enters the prediction of the width
  difference $\dg_s$, for which we obtain $\dg_s^{\rm SM} = 0.088 \pm
  0.017 \, \mbox{ps}^{-1}$, if no new physics enters \bbms. Applying
  our formulae to Tevatron data we find a deviation of the \bbms\ phase 
  $\phi_s$ from its Standard Model value by 2 standard deviations.   
  I stress that present data do not give any information on the 
  sign of $\dg_s$. 
\end{abstract}

\ccode{PACS numbers: 12.38.Bx, 13.25.Hw, 11.30Er, 12.60.-i}

\section{Introduction}    
Flavor-changing neutral current (FCNC) processes of quarks are highly
sensitive to new physics at or above the electroweak scale. The highly
successful physics program of the B factories has revealed that these
FCNCs are dominantly governed by the Cabibbo-Kobayashi-Maskawa (CKM)
mechanism\cite{ckm} of the Standard Model. This is an important
constraint on the new, still undiscovered theory of Tera--scale physics.
Models whose only source of flavor violation is the CKM matrix are
termed \emph{minimally flavor violating (MFV)}\cite{dgis}. Still, it is
easy to construct models in which non--MFV are naturally suppressed
while still leading to measurable effects. An example are Grand Unified
Theories (GUTs) in which MFV is implemented far above the GUT scale, but
subsequently altered by renormalisation group effects\cite{cmm}.
Particularly well-suited for the search of corrections to the CKM
mechanism are CP--violating observables in $b\to s$ transitions:
CKM--driven CP violation in $b\to s$ transitions is small, just as in
the $s\to d $ transitions probed in Kaon physics. Interestingly, in
supersymmetric GUT models the large atmospheric neutrino mixing angle
can influence $b\to s$ transitions\cite{cmm,jn}. Clearly, the ``holy
grail'' of $b\to s$ FCNC physics is the \bbms\ amplitude $M_{12}^s$.
New physics will affect its magnitude and phase, and already small
contributions to $M_{12}^s$ can lift the small (and in many cases
unobservably tiny) CP asymmetries to sizable values. 

In this talk I summarize the avenues to (over--)constrain the \bbms\ 
complex. Then a new, more precise, theory prediction for the element
$\Gamma_{12}^s$ of the decay matrix is presented. $\Gamma_{12}^s$ enters
the formulae for the width difference $\dg_s$ of the two mass
eigenstates in the $B_s$ system. Our new result further decreases the
theoretical uncertainty in the extraction of the  \bbms\ phase 
from the CP asymmetry in flavor--specific $B_s$ decays. Finally,
constraints from D\O\ data on $M_{12}^s$ are derived with the help of
the new result for $\Gamma_{12}^s$. The presented results are from
Ref.~\cite{ln}. 
 
\section{\bbms}    
\bbs\ oscillations are governed by a Schr\"odinger
equation
\begin{equation}
i \frac{d}{dt}
\left(
\begin{array}{c}
\ket{B_s(t)} \\ \ket{\bar{B}_s (t)}
\end{array}
\right)
=
\left( M^s - \frac{i}{2} \Gamma^s \right)
\left(
\begin{array}{c}
\ket{B_s(t)} \\ \ket{\bar{B}_s (t)}
\end{array}
\right)\label{sch}
\end{equation}
with the mass matrix $M^s$ and the decay matrix $\Gamma^s$.  The
physical eigenstates $\ket{B_H}$ and $\ket{B_L}$ with the masses
$M_H,\,M_L$ and the decay rates $\Gamma_H,\,\Gamma_L$ are obtained by
diagonalizing $M^s-i \Gamma^s/2$.  There are three 
physical quantities in \bbms: 
\begin{eqnarray}
\quad&& |\Gamma_{12}^s|, \qquad  |M_{12}^s|\quad\mbox{and }\quad
 \phi_s=\arg(-M_{12}^s/\Gamma_{12}^s) . \label{pq}
\end{eqnarray}
The phase $\phi_s$ is responsible for \emph{CP violation in mixing}.
The mass and width differences between $B_{L}$ and $B_{H}$ are
\begin{eqnarray}
\dm_s &\equiv & M^s_H -M^s_L \; = \; 2\, |M_{12}^s|, \nn
\dg_s & \equiv & \Gamma^s_L-\Gamma^s_H \; =\; 
     2\, \real \frac{\Gamma_{12}^s}{M_{12}^s} \; =\; 
        2\, |\Gamma_{12}^s| \cos \phi_s, \label{dmdg}
\end{eqnarray}
Here and in the following I neglect numerically irrelevant corrections
of order $m_b^2/M_W^2$. The average width of the two eigenstates is
denoted by $\Gamma_s=(\Gamma_L+\Gamma_H)/2$.  The precise measurements
from the D\O\ and CDF experiments\cite{ichep06dm}
\begin{eqnarray}
17 \, \mbox{ps}^{-1} \leq  \dm_s & \leq & 21 \, \mbox{ps}^{-1}
  \qquad\qquad \quad @90\% \, \mbox{CL}  \quad\qquad  \mbox{D\O} \nn
\dm_s & =&  17.77\pm{0.10} {}_{\mbox{\scriptsize (syst)}}
       \pm 0.07\,{}_{\mbox{\scriptsize (stat)}} \, \mbox{ps}^{-1} \,
   \qquad \mbox{CDF} . \label{dmexp}
\end{eqnarray}
determine $|M_{12}^s|$ sharply. 

New physics affects magnitude and phase of $M_{12}^s$, but can barely
change $\Gamma_{12}^s$, which dominantly stems from the Cabibbo-favored
tree--level $b\to c\ov c s$ decays. In the Standard Model $\phi_s$ is
tiny, so that $\cos\phi_s\simeq 0$ and new physics can only decrease
$\dg_s$ in \eq{dmdg}\cite{g,dfn}.  The success of the Standard Model
suggests that new physics enters low--energy observables at the loop
level. A different viewpoint has recently been taken in Ref.~\cite{dkn},
where potentially huge contributions to $\Gamma_{12}^s$ from leptoquarks
with FCNC couplings have been claimed. The authors of Ref.~\cite{dkn}
exploit the plethora of free parameters in leptoquark models to place
their effect into the yet unmeasured decay $B_s \to \tau^+\tau^-$, whose
branching ratio could then be enhanced to up to 18\%.  The effect of
some new decay mode on the ratio $\dg_s/\Gamma_s$ can be as large as
twice its branching ratio, so that Ref.~\cite{dkn} finds an enhancement
of $\dg_s/\Gamma_s$ from its Standard Model value around 0.15 to up to
0.51. However, the enhanced $\ov s b \ov \tau \tau$ coupling invoked in
this model would also lead to sizable new $b\to s \tau^+ \tau^-$ decays
modes of $B^+$ and $B_d$ mesons. While there are no precise data on
final states with two $\tau$'s yet, the extra decay modes would sizably
alter well-measured and well-calculated inclusive quantities such as the
semileptonic branching ratio $B_{\rm SL}$\cite{bbbg}, the inclusive
branching ratio ${\cal B} (B\to \mbox{ no charm})$\cite{gln} and even the
lifetimes of all B mesons. Thus the idea of Ref.~\cite{dkn} is not
viable and it is safe to assume that $\Gamma_{12}^s$ is unaffected by
new physics.

While the pristine measurement in \eq{dmexp} already gives a powerful
constraint on new physics\cite{}, there are two reasons to seek further
experimental information from other observables: first, $\dm_s$ only
constrains $|M_{12}^s|$ but not $\phi_s$ and second,  the translation 
of \eq{dmexp} into constraints on fundamental parameters involves a
hadronic parameter, which is difficult to compute and inflicts a
theoretical uncertainty of order 30\% onto the analysis. New physics 
entering $M_{12}^s$ can be parameterized as 
\begin{eqnarray}
M_{12}^s & \equiv & M_{12}^{s, \rm SM} \cdot  \Delta_s \, ,
\qquad\qquad  \Delta_s \; \equiv \;  |\Delta_s| e^{i \phi^\Delta_s} .
 \label{defdel}
\end{eqnarray}
Thus every measurement related to \bbms\ gives a constraint on the
complex $\Delta_s$ plane. The Standard Model corresponds to
$\Delta_s=1$. While $\arg M_{12}^s$ is unphysical and depends on phase
conventions, $\phi^\Delta_s$ is a physical CP phase. The mixing phase
$\phi_s $ in \eq{pq} can be written as 
\begin{eqnarray}
 \phi_s &=& \phi^\Delta_s \; +\; \phi_s^{\rm SM}, \no
\end{eqnarray}   
where $\phi_s^{\rm SM}$ is the Standard Model prediction for $\phi_s$. 
$\phi_s^{\rm SM}$ is negligible, see \eq{finphi} below. 

The relationship of $\Delta_s$ to the parameters used in
\cite{nir2006,exput} is
\begin{eqnarray}
\Delta_s &=& r_s^2 e^{2 i \theta_s}. \no
\end{eqnarray}
We find it more transparent to plot $\imag \Delta_s$ vs.\ $\real
\Delta_s$ than to plot $2\theta_s$ vs.\ $r_s^2$.  Next we list the key
measurements which (over--)constrain $\Delta_s$:
\begin{itemize}
\item[1)] $\dm_s$ in \eq{dmexp} determines $|\Delta_s|$. 
\item[2)] Measuring the lifetime in an untagged $b\to c\ov c
  s$ decay $\BsorBsbar \to f_{CP}$, where $f_{CP}$ is a CP eigenstate,
  determines $\dg_s \cos (\phi_s^\Delta-2\beta_s)= |\dg_s \cos
  (\phi_s^\Delta-2\beta_s)|$\cite{g,dfn}.  
  The time-dependent decay rate reads
\begin{eqnarray}
\Gamma[ \BsorBsbar \to f_{CP\pm},t ] &\propto&
        \frac{1 \pm \cos(\phi_s^\Delta -2 \beta_s) }{2} e^{-\Gamma_L t}
    \; +\:
         \frac{1\mp \cos(\phi_s^\Delta -2 \beta_s) }{2} e^{-\Gamma_H t} \nn
& =& e^{-\Gamma_s t} \lt[ \cosh \frac{\dg_s \, t}{2} \mp 
       \cos(\phi_s^\Delta -2 \beta_s) \sinh \frac{\dg_s \, t}{2}
       \rt]. 
\label{twoexp}
\end{eqnarray}
Here the sign convention of
\begin{eqnarray}
\beta_s & = & - \arg \lt( - \frac{\lambda_t^s}{\lambda_c^s}\rt)
   \; =\; 0.020 \pm 0.005 \; = \; 1.1^\circ \pm 0.3^\circ 
\label{defbetas}
\end{eqnarray}
is that of Ref.~\cite{run2}. Currently this measurement is applied to
$\BsorBsbar \to J/\psi \phi$. Here the CP quantum number of the final
state depends on the orbital angular momentum, the P--wave state is
$f_{CP-}$ and the S--wave and D--wave components correspond to $f_{CP+}$
in \eq{twoexp}.  Neglecting $\beta_s$, $\phi_s^{\rm SM}$ and expanding
to first order in $\dg_s$ one verifies from \eq{twoexp} that the
lifetime measurement determines\cite{g,dfn}
\begin{eqnarray}
\dg_s \cos \phi_s^\Delta &=& 2 |\Gamma_{12}^s| \cos^2 \phi_s^\Delta. 
\label{gc2}
\end{eqnarray}
\item[3)] The angular analysis of an untagged $\BsorBsbar \to J/\psi
  \phi$ sample not only determines $\dg_s \cos \phi_s^\Delta $ 
  as discussed in item 2, but also contains information on
  $\sin (\phi_s^\Delta -2\beta_s)$ through a CP-odd interference term.
\item[4)] The CP asymmetry in \emph{flavor-specific} $B_s\to f$
decays is 
\begin{eqnarray}
a^s_{\rm fs}
     &=&
    \imag \frac{\Gamma_{12}^s}{M_{12}^s}
    \; = \; \frac{|\Gamma_{12}^s|}{|M_{12}^s|} \sin \phi_s
    \; = \; \frac{\dg_s}{\dm_s} \tan \phi_s
 . \label{defafs}
\end{eqnarray}
$a_{\rm fs}^s$ is typically measured by counting the number of
positively and negatively charged leptons in semileptonic $\BsorBsbar$
decays.  Observing further the time evolution of these untagged
$\BsorBsbar \to X^\mp \ell^\pm \nuornubar_\ell$ decays,
\begin{eqnarray}
\hspace{-6ex}
\frac{ \Gamma[ \BsorBsbar \to X^- \ell^+\nu_\ell ,t ] \, -\,
       \Gamma[ \BsorBsbar \to X^+ \ell^- \ov \nu_\ell ,t ]}{
       \Gamma[ \BsorBsbar \to X^- \ell^+\nu_\ell ,t ] \, +\,
       \Gamma[ \BsorBsbar \to X^+ \ell^- \ov \nu_\ell ,t ]}
&=& \frac{a_{\rm fs}^s}{2} \lt[ 1-
    \frac{\cos(\dm_s \, t)}{\cosh{(\dg_s\, t/2)}}  \rt],
\label{afst}
\end{eqnarray}
may help to control systematic experimental effects\cite{n}. 
\end{itemize}
Often the average lifetime of the two $B_s$ eigenstates is included into
global experimental analyses of $\dg_s$. Heavy quark physics implies
that the average widths $\Gamma_d$ and  $\Gamma_s$ in the $B_d$ and
$B_s$ systems are equal up to corrections of order 1\%. The average
$B_s$ lifetime will then exceed the $B_d$ lifetime by a term which is
quadratic in $\dg_s^2/(\Gamma_s^2)$\cite{g,dfn}. For realistic values of
$\dg_s$ this term is too small to determine $\dg_s$. 

We close this section with an important remark: up to now there is
\emph{no} experimental information on the sign of $\dg_s$ available!
From \eq{dmdg} one verifies immediately that $\sgn \dg_s= \sgn
\cos\phi_s$ and (see \eq{gc2}) the untagged analysis described in item 2
is only sensitive to $|\dg_s|$. Also the CP--odd quantities described in
items 3 and 4 only determine $\sin\phi_s^\Delta$, which comes with a
two--fold ambiguity for $\phi_s^\Delta$: the two solutions correspond to
different signs of $\cos\phi_s^\Delta$ and thereby different signs of
$\dg_s$. The determination of $\sgn \dg_s$ is discussed in
Ref.~\cite{dfn}. It is easy to check that the formula for the angular
distribution in $\BsorBsbar \to J/\psi \phi$\cite{ddlr,dfn} is unchanged if
one simultaneously flips the sign of $\dg_s$ (i.e.\ interchanges
$\Gamma_L$ and $\Gamma_H$) and the sign of $\cos \phi_s$. Thus current
experimental results should be quoted for $|\dg_s|$ rather than $\dg_s$.

\section{New theory prediction for $\Gamma_{12}^s$} 
The predictions of $M_{12}^s$ and $\Gamma_{12}^s $ involve hadronic
matrix elements of four--quark operators. These matrix elements are
computed with the help of lattice gauge theory and dominate the
theoretical uncertainty. In the Standard Model prediction for $M_{12}^s$
one only encounters the operator
\begin{eqnarray}
Q & =&   \ov s_\alpha \gamma_\mu (1-\gamma_5) b_\alpha \,
         \ov s_\beta \gamma^\mu (1-\gamma_5) b_\beta ,
\label{defq}
\end{eqnarray}
where $\alpha$ and $\beta$ are color indices.  The matrix element is
usually parameterized as
\begin{eqnarray}
\bra{B_s} Q \ket{\ov B_s}  &=& \frac{8}{3} M^2_{B_s}\, f^2_{B_s} B
      , \label{defb}
\end{eqnarray}
where $M_{B_s}$ and $f_{B_s}$ are mass and decay constant of the $B_s$,
respectively. $B$ is called a bag factor. Then\cite{bjw} 
\begin{eqnarray}
\dm_s^{\rm SM} & = & \lt( 19.3 \pm 0.6 \rt) \, \mbox{ps}^{-1}
          \lt( \frac{|V_{ts}|}{0.0405}\rt)^2  \cdot
          \lt( \frac{f_{B_s}}{240 \, \mbox{MeV}} \rt)^2
             \frac{B}{0.85} \label{dms} \\
  & = & \lt( 19.30 \pm 6.68 \rt) \, \mbox{ps}^{-1}
\label{dmsnum} .
\end{eqnarray}
The number in \eq{dmsnum} is found from \eq{dms} with $f_{B_s}=240 \pm
40 \, \mbox{MeV}$ and $B = 0.85 \pm 0.06$\cite{lattice,sum}. 

The situation with $\Gamma_{12}^s$ is more complicated: its prediction
requires the expansion in two parameters, $\lbar/m_b$ and
$\alpha_s(m_b)$. Here $\lbar \sim (M_{B_s}-m_b)$ is the relevant
hadronic scale and $\alpha_s$ is the strong coupling
constant\cite{bbd1,bbgln1,bbln,rome03}. In the first step one finds that
three operators contribute to $\Gamma_{12}^s$ at leading order in
  $\lbar/m_b$: $Q$ defined in \eq{defq}, 
\begin{eqnarray}
Q_S & = &  \ov{s}_\alpha (1+\gamma_5)  b_\alpha \,
           \ov{s}_\beta  (1+\gamma_5)  b_\beta 
  \label{defqs} \\
\mbox{and}\qquad \widetilde{Q}_S & = &  
     \ov{s}_\alpha (1+\gamma_5)  b_\beta \,
                       \ov{s}_\beta (1+\gamma_5)  b_\alpha,
  \label{defqst}
\end{eqnarray}
We parameterize the new matrix elements with bag factors 
$B_S^\prime$ and $\widetilde B_S^\prime$.
Subsequently one trades one of the operators for 
\begin{eqnarray}
R_0 &\equiv&                            Q_S
              \; + \; \alpha_1  \tilde Q_S
           \; + \; \frac{1}{2}  \alpha_2 Q,
\label{defr0}
\end{eqnarray}
Here $\alpha_{1,2}=1 +{\cal O}(\alpha_s(m_b))$ are QCD correction
factors\cite{bbgln1,ln}. This is done, because the
matrix element of $R_0$ is suppressed by $\lbar/m_b$, so that it belongs
to the subleading order\cite{bbd1}. In Refs.~\cite{bbd1,bbgln1,rome03}
\eq{defr0} has been used to eliminate $\widetilde{Q}_S$ from the
operator basis.  This results in the prediction
\begin{eqnarray}
\Delta \Gamma_{s,\rm old}^{\rm SM} & = & 
2 |\Gamma_{12, \rm old }^s| \; =\; 
\left( \frac{f_{B_s}}{240 \, \mbox{MeV}} \right)^2
\left[ 0.002 B + 0.094  B_S^\prime - \rt. \nn
&& \qquad\qquad\qquad\qquad \lt.
\left(0.033  B_{\tilde{R}_2} +  0.019 B_{R_0}+  0.005 B_R \right) \right]
\, \mbox{ps}^{-1}, \label{oldr} 
\end{eqnarray}
where $B_{R_0}$ and $B_{\tilde{R}_2}$ are the bag factors of $R_0$ and
another subleading operator, $\tilde{R}_2$, and the uncertainties of the
coefficients are not shown. The other sub-leading operators come with
smaller coefficients and are accounted for with a common bag factor $B_R$ in
\eq{oldr}. This result is pathological in several respects: the
$\lbar/m_b$ corrections exceed their natural size of 20\% and are
negative, the next--to--leading order QCD corrections of
Refs.~\cite{bbgln1,rome03} (which are contained in the numbers 0.002 and
0.094) are also large and decrease the result further and finally
$\Delta \Gamma_{s,\rm old}^{\rm SM}$ is dominated by $B_S^\prime$, so
that the cancellation of hadronic physics from the ratio $\dm_s/\dg_s$
is imperfect.

The starting point of the improvement in Ref.~\cite{ln} is the
observation that the matrix element of $\widetilde Q_S$ is
small\cite{bgmpr}.  Keeping in mind that the matrix element of $R_0$ is
power--suppressed and therefore also small, \eq{defr0} encodes a strong
numerical correlation between $B$ and $B_S^\prime$. \eq{defr0} implies
for the bag parameters:
\begin{eqnarray}
\alpha_1  \widetilde B_S^\prime - 5 B_S^\prime
           \; + \; 4  \alpha_2 B &=& {\cal O} 
        \lt( \frac{\lbar}{m_b} \rt).  
\label{relb}
\end{eqnarray}
Hence trading $\widetilde B_S^\prime$ for a linear combination of $B$,
$B_S^\prime$ and $B_{R_0}$ expresses a small number in terms of the
difference of two big numbers: $\widetilde B_S^\prime=5 B_S^\prime - 4 B
+ {\cal O} (\lbar/m_b,\alpha_s )$. So one tends to introduce a
theoretical uncertainty into the problem, which is not inherent to the
calculated quantity. The most straightforward way to take care of this
is to keep $Q$ and $\widetilde Q_S$ in the basis and to abandon $Q_S$
instead. This results in\cite{ln}
\begin{eqnarray}
\Delta \Gamma_s^{\rm SM} 
& = & \left( \frac{f_{B_s}}{240 \, \mbox{MeV}} \right)^2
\Big[ (0.105 \pm 0.016) B + (0.024 \pm 0.004)  \tilde B_S^\prime 
\nonumber
\\
&&
\hspace{0.5cm} -
\left[ (0.030 \pm 0.004) B_{\tilde{R}_2} -
       (0.006\pm 0.001) B_{R_0} +
        0.003 B_R \right] \!\Big] \, \mbox{ps}^{-1}. \quad
\label{finaldg} 
\end{eqnarray}
The quoted result further includes the resummation of logarithms of the
charm mass to all orders in perturbation theory.  Now all pathologies
have disappeared. The smallness of the coefficient of $B_{R_0}$ compared
to \eq{oldr} can be understood with the help of the $1/N_c$ expansion.
($N_c=3$ is the number of colors.) The coefficients of $Q$ and
$\widetilde Q_S$ are leading in $1/N_c$, while the coefficient of $Q_S$
is color--suppressed. If one eliminates $Q_S$ in terms of $R_0$, the
coefficient of $Q_S$ becomes the coefficient of $R_0$, so that the
number multiplying $ B_{R_0}$ is small. In the old result in \eq{oldr},
however, the coefficient of $ B_{R_0}$ stems from the color-favored
coefficient of $\widetilde Q_S$ and is large. On the other hand, the
(equally welcome) reduction of the NLO QCD correction, which is related to
the QCD factors $\alpha_{1,2}$, appears accidental.

From our new result for $\Gamma_{12}^s$  we also get a new prediction
for $a_{\rm fs}^s$ and the CP phase $\phi_s$ in the Standard Model. 
Our predictions are 
\begin{eqnarray}
\Delta \Gamma_s^{\rm SM} & = & 
\left( 0.096   \pm 0.039 \right) \, \mbox{ps}^{-1}
 \hspace{0.15cm} \Rightarrow \hspace{0.15cm}
\frac{\Delta \Gamma_s^{\rm SM}}{\Gamma_s}  = \Delta \Gamma_s^{\rm SM} \cdot
 \tau_{B_d} = 0.147 \pm 0.060 \; \; \label{findg}
\\
a_{\rm fs}^{s, \rm SM} & = & \left( 2.06 \pm 0.57 \right) \cdot 10^{-5}
\\
\frac{\Delta \Gamma_s^{\rm SM}}{\Delta M_s^{\rm SM}}  & = &
\left( 49.7 \pm 9.4 \right) \cdot 10^{-4} \label{dgdmr}
\\
\phi_s^{\rm SM} & = & (4.2\pm 1.4 )\cdot 10^{-3}
      \; = \; 0.24^\circ \pm 0.08^\circ
\label{finphi}
\end{eqnarray}
For the precise values of the input parameters I refer to
Ref.~\cite{ln}.

The prediction of the ratio $\dg_s/\dm_s$ in \eq{dgdmr} got much sharper,
because most of the hadronic uncertainies now cancel, since $\dg_s$ in
\eq{finaldg} is dominated by the term involving $B$.  With \eq{dmexp}
one finds from \eq{dgdmr}:
\begin{eqnarray}
\Delta \Gamma_s^{\rm SM} & = &
\frac{\Delta \Gamma_s^{\rm SM}}{\Delta M_s^{\rm SM}} 
\cdot \Delta M_s^{\rm exp} = 0.088 \pm 0.017
\, \mbox{ps}^{-1}
\\
& \Rightarrow &
\frac{\Delta \Gamma_s^{SM}}{\Gamma_s}
= \Delta \Gamma_s^{\rm SM} \cdot \tau_{B_d} = 0.127 \pm 0.024 \, .
\end{eqnarray}  
Any future measurement of $\dg_s$ outside this range will signal
new physics in $\dm_s$ or $\dg_s$.  

For predictions of the corresponding quantities in the \bbmd\ complex I
refer to Refs.~\cite{ln,n2}.  

\section{Constraining new physics} 
The formulae relating $\dm_s$, $\dg_s$ and $a_{\rm fs}^s$ to $\Delta_s$
defined in \eq{defdel} are
\begin{eqnarray}
\dm_s  & = & \dm_s^{\rm SM} \,  |\Delta_s|
=
(19.30 \pm 6.74 ) \, \mbox{ps}^{-1} \cdot | \Delta_s|
\label{bounddm}
\\
\Delta \Gamma_s  & = & 2 |\Gamma_{12}^s|
     \, \cos \left( \phi_s^{\rm SM} + \phi^\Delta_s \right)
= (0.096 \pm 0.039) \, \mbox{ps}^{-1}
\cdot \cos \left( \phi_s^{\rm SM} + \phi^\Delta_s \right)
\label{bounddg}
\\
\frac{\Delta \Gamma_s}{\Delta M_s}
&= &
 \frac{|\Gamma_{12}^s|}{|M_{12}^{\rm SM,s}|}
\cdot \frac{\cos \left( \phi_s^{\rm SM} + \phi^\Delta_s \right)}{|\Delta_s|}
=
\left( 4.97 \pm 0.94 \right) \cdot 10^{-3}
\cdot \frac{\cos \left( \phi_s^{\rm SM} + \phi^\Delta_s \right)}{|\Delta_s|}
\label{bounddgdm}
\\
a_{\rm fs}^s
&= &
 \frac{|\Gamma_{12}^s|}{|M_{12}^{\rm SM,s}|}
\cdot \frac{\sin \left( \phi_s^{\rm SM} + \phi^\Delta_s \right)}{|\Delta_s|}
= \left( 4.97 \pm 0.94 \right) \cdot 10^{-3}
\cdot \frac{\sin \left( \phi_s^{\rm SM} + \phi^\Delta_s
  \right)}{|\Delta_s|}
\label{boundafs} 
\end{eqnarray}
For our analysis we use the the CDF data on $\dm_s$ in \eq{dmexp} and
D\O\ data on the angular distribution in $\BsorBsbar \to J/\psi
\phi$\cite{dgexpnew}, the semileptonic CP asymmetry $a_{\rm sl}^s=a_{\rm
  fs}^s$\cite{aslsexp} and on the same--sign di--muon asymmetry $a_{\rm
  sl}$\cite{dimuonexp}, which is related to $a_{\rm sl}^s$ and $a_{\rm sl}^d$
as (updated from Ref.~\cite{nir2006})
\begin{equation}
a_{\rm sl} = \left(0.582 \pm 0.030 \right) \, a_{\rm sl}^d +
         \left(0.418 \pm 0.047 \right) \, a_{\rm sl}^s .
\label{asllk}
\end{equation}
The angular analysis of $\BsorBsbar \to J/\psi \phi$\cite{ddlr,dfn}
involves two strong phases $\delta_1$ and $\delta_2$. They can be
determined from the data, albeit with discrete ambiguities which imply a
four--fold ambiguity in $\phi_s$. The CP--conserving piece of the
angular distribution depends on $\cos (\delta_2-\delta_1)$, so that the
experimental error on $\cos (\delta_2-\delta_1)$ is smaller than the
error on $\cos\delta_1$ and $\cos\delta_2$, which appear in the
CP--violating piece proportional to $\sin(\phi_s^\Delta-2\beta_s)$. 
The D\O\ result for $
\delta_2-\delta_1=\pm(2.6\pm0.4)$\cite{dgexpnew} is in good agreement 
with theoretical model calculations predicting $\delta_1 \sim \pi$ and
$\delta_2\sim 0$\cite{bsw}. In our analysis we have fixed 
$\cos\delta_1 <0 $ and $\cos\delta_2 >0 $, which reduces the four--fold
ambiguity in $\phi_s^\Delta$ to a two--fold one. D\O\ finds\cite{dgexpnew}
\begin{eqnarray}
\Delta \Gamma_s & = & \phantom{-}
       0.17 \pm 0.09{}_{\mbox{\scriptsize (stat)}} \;
          \pm 0.03{}_{\mbox{\scriptsize (syst)}}  \; \mbox{ps}^{-1} \,
\nn &&
\mbox{and} \quad
\phi_s^\Delta-2\beta_s 
\; = \; -0.79  \pm 0.56{}_{\mbox{\scriptsize (stat)}}
           \pm 0.01{}_{\mbox{\scriptsize (syst)}}  \label{eqdgexp5}\\[2mm]
\mbox{or}\qquad
\Delta \Gamma_s & = & - 0.17 \pm 0.09{}_{\mbox{\scriptsize (stat)}}
    \pm 0.03{}_{\mbox{\scriptsize (syst)}}  \, \mbox{ps}^{-1} \,
\nn &&
\mbox{and} \quad
\phi_s^\Delta-2\beta_s 
\; = \; -0.79  \pm 0.56{}_{\mbox{\scriptsize (stat)}}
           \pm 0.01 {}_{\mbox{\scriptsize (syst)}} \; +\; \pi  \, .
\label{eqdgexp4}
\end{eqnarray}
The second solution deviates from the Standard Model by several standard
deviations. In order to assess the compatibility of the measurement with
the Standard Model we only need to consider the first solution in
\eq{eqdgexp5}. 

\begin{figure}[t]
  \centerline{\psfig{file=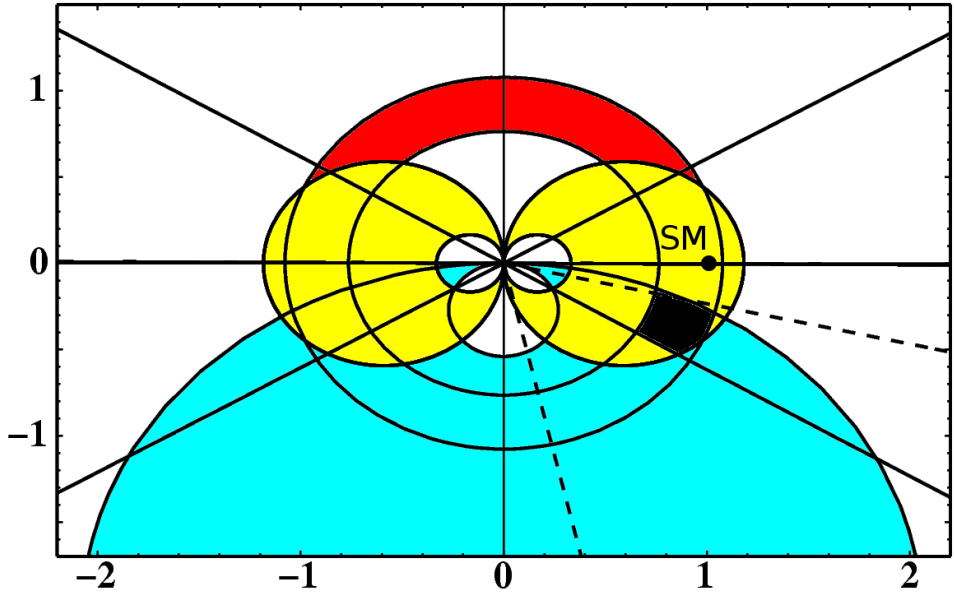,width=\textwidth}} \vspace*{8pt}
\caption{Current experimental bounds in the complex $\Delta_s$-plane.
  The bound from $\Delta M_s$ is the red (dark-grey) annulus around the
  origin. The bound from $|\Delta \Gamma_s|/ \Delta M_s$ corresponds to
  the yellow (light-grey) region and the bound from $a_{\rm fs}^s$ is
  given by the light-blue (grey) region. The angle $\phi_s^\Delta$ can
  be extracted from $|\Delta \Gamma_s|$ (solid lines) with a four--fold
  ambiguity --- each of the four regions is bounded by a solid ray and
  the x-axis --- or from the angular analysis in $B^0_s \to J / \psi \phi$
  (dashed line). (No mirror solutions from discrete ambiguities are
  shown for the latter.)  The current experimental situation shows a
  $2\sigma$ deviation from the Standard Model case $\Delta_s=
  1$.\label{boundbandreal}}
\end{figure}

The same--sign di--muon asymmetry $a_{\rm sl}$ in
\eq{asllk} involves both $a_{\rm sl}^s$ and $a_{\rm sl}^d$. In order to
translate the D\O\ measurement\cite{dimuonexp} of 
\begin{equation}
a_{\rm sl} = \left( -2.8 \pm 1.3 {}_{\mbox{\scriptsize (stat)}}
                         \pm 0.9 {}_{\mbox{\scriptsize (syst)}}
             \right) \cdot 10^{-3} \,
\label{eqdimuexp}
\end{equation}
into a number for $a_{\rm sl}^s$ we have used our theory prediction for 
 $a_{\rm sl}^d$\cite{bbln,ln}. Combining the result with the measurement 
of $a_{\rm sl}^s$\cite{aslsexp} gives 
\begin{equation}
a_{\rm sl}^{s} = \left(- 5.2 \pm 3.2{}_{\mbox{\scriptsize (stat)}}
                             \pm 2.2{}_{\mbox{\scriptsize (syst)}}
                 \right) \cdot 10^{-3} \, .
\label{eqasldfinal}
\end{equation}
\fig{boundbandreal} shows the combination of all measurements. 
It indicates a deviation from the Standard
Model value $\Delta_s=1$ by $2\sigma$.  

For this conclusion it is crucial that we use the theoretical value for
$a_{\rm fs}^d$\cite{ln}, which is much more precise than the current
measurement of this quantity (see Ref.~[14] of Ref.~\cite{cdz}):
\begin{eqnarray}
a_{\rm fs}^{d,\rm th} &=& \lt(  -4.8\epm{1.0}{1.2} \rt) \,
                \cdot 10^{-4}, \qquad \quad
a_{\rm fs}^{d,\rm exp} \; =\; \lt( -47 \pm 46\rt)  \cdot 10^{-4}
.  \label{afsnum} 
\end{eqnarray}
This assumes, of course, that no new physics enters $a_{\rm fs}^d$.
Another combined analysis of the D\O\ data of
Refs.~\cite{dgexpnew,aslsexp,dimuonexp} has been performed in
Ref.~\cite{cdz}.  This analysis has used the experimental value $a_{\rm
  fs}^{d,\rm exp}$ in \eq{asllk}, which leads to a significantly weaker
constraint from the same--sign di--muon asymmetry. Also no theory input
on $2|\Gamma_{12}^s|$ has been used in the analysis of Ref.~\cite{cdz}.
The final result for $\phi_s$ is therefore more conservative than ours
and implies a deviation from the Standard Model by only $1.5\sigma$.

\section{Conclusions}
I presented an improved theoretical prediction of $\Gamma_{12}^s$, which
permits more accurate predictions of the width difference $\dg_s$ and of
the CP asymmetry in flavor-specific decays, $a_{\rm fs}^s$, in
scenarios of physics beyond the Standard Model.  Applying the new
formulae to D\O\ data we find that the \bbms\ phase deviates from the
Standard Model value by $2\sigma$. We conclude that current experiments
are reaching the sensitivity to probe new physics in the \bbms\ phase.

\section*{Acknowledgments}
I acknowledge the warm hospitality of the organizers of the CTP07
workshop. 

This work was supported in part by the DFG grant No.~NI 1105/1--1, by the
EU Marie-Curie grant MIRG--CT--2005--029152, by the BMBF grant 05 HT6VKB
and by the EU Contract No.~MRTN-CT-2006-035482, \lq\lq FLAVIAnet''.

\appendix

\end{document}